\documentstyle[12pt]{article}
\begin{document}
\setcounter{page}{0}
\thispagestyle{empty}

\begin{center}

\vfill

{\Large \bf Detailed analysis of the dependence of the one-loop
counterterms on the gauge and parametrization in the Einstein gravity
with the cosmological constant.}
\vspace{1cm}

{\large M.~Yu.~Kalmykov}
\footnote { E-mail: $kalmykov@thsun1.jinr.dubna.su$},

\vspace{1cm}

{\em Bogoliubov Laboratory of Theoretical Physics,
 Joint Institute for Nuclear  Research,
 $141~980$ Dubna $($Moscow Region$)$, Russian Federation}

\vspace{1cm}

{\large K.~A.~Kazakov}
\footnote{E-mail: $kirill@theor.phys.msu.su$},
~~{\large P.~I.~Pronin}
\footnote{E-mail: $petr@theor.phys.msu.su$},
~~{\large K.~V.~Stepanyantz}
\footnote{E-mail: $stepan@theor.phys.msu.su$}.

\vspace{1cm}

{\em Moscow State University, Physical Faculty,\\
Department of Theoretical Physics.\\
$117234$, Moscow, Russia}
\end{center}
\vfill

\begin{abstract}
In this paper, the dependence of the Einstein gravity with the
cosmological constant as well as of this theory in the first-order
formalism on the gauge and parametrization is been analyzed.
The one-loop counterterms off the mass shell have been plainly
calculated in arbitrary gauge and parametrization. The tensor package
of analytic calculations, written in REDUCE, allowed all the
calculations to be carried out.  A method of renormalization
group functions calculations off shell is discussed.
\end{abstract}


\pagebreak
\section{Introduction}

At present, there is no gravity which is both perturbatively
renormalizable and unitary. The Einstein theory is finite in the
one-loop approximation on the motion equations \cite{TV}, however,
non-renormalizable at two-loops \cite{GS,VDV,VT1}. Moreover, the
interaction with matter fields leads to counterterms which are
non-renormalizable even at one-loop \cite{matter}. The supergravity
is unproven to be renormalizable at an arbitrary order of
perturbation theory \cite{SUGRA}. The theory, which is quadratic in
the curvature, is renormalizable \cite{st,VT2} and has the
asymptotical freedom \cite{asymp}, but this theory is not
unitary in the linear approximation \cite{BOS}. All attempts to
restore the unitarity at the expense of either quantum corrections or
an interaction with matter fields failed \cite{unitarity}. Hopefully,
this problem can be solved in the framework of
non-perturbative approaches \cite{nonpertur}.  The superstring
approach to the quantum gravity \cite{GSW} and the canonical one
\cite{A1} recently proposed are claimed as "sensible" theories of
quantum gravity.

It is commonly believed that the quantum gravitational corrections
can drastically change the classical behaviour of the theory. Due to
these quantum corrections it becomes possible to avoid appearance of
singularities, to modify the Newtonian potential, to solve the
problem of cosmological constant smallness, etc.  Since there is no
reliable quantum gravity, one should deal with effective models.
Among them, the Einstein theory and the $R^2$-gravity are the most
investigated and usable. Quantum properties of these theories are
widely explored. However, in our opinion, the dependence of the
quantum gravity on the parametrization was not studied enough.
We consider parametrization arbitrariness as an arbitrariness in choice
of dynamical variables in gravity. Both the tensor densities
$g^{\ast}_{\mu \nu } = g_{\mu \nu }(-g)^r$ and $g^{\ast \mu \nu } =
g^{\mu \nu }(-g)^s$ can be considered as a gravitational field.
The number of interaction vertices which appear at calculation of
radiative corrections and their types can be significantly reduced
choosing the appropriate gravitational variables. For example,
if we consider $g_{\mu \nu }$ as a gravitational variable, the number
of three-point interactions in the Einstein gravity is equal to
13~\cite{GS}; if the tensor density $g^{\mu \nu } \sqrt{-g}$ is
selected as a dynamical variable, the number of three-point
interactions is equal to 6~\cite{CLR}.  The theorem of equivalence
\cite{equiv} says that the $S$-matrix of renormalizable theories is
invariant under the following replacement:

$$
\varphi^j \rightarrow '\varphi^j = \varphi^j +
\left( \varphi^2 \right)^j + \left( \varphi^3 \right)^j + \dots .
$$

\noindent
It was shown for an arbitrary gauge theory that the effective
action (its finite part and divergencies independently) on the mass shell
does not depend on the gauge and parametrization \cite{VLT}. Whereas the
off-shell Green's functions are dependent on the gauge and
parametrization.

So, for getting the physical quantities off the mass shell we should
get rid of the evident parametrization dependence. It can be done in
two ways.  First of them is related to using the Vilkovisky-DeWitt
effective action which is independent of the gauge and
parametrization instead of the standard effective action \cite{VDW}.
But the calculations of loop corrections to the Vilkovisky-DeWitt
effective action are very cumbersome due to nonlocal terms in its
definition. Moreover, there is a dependence of the
Vilkovisky-DeWitt effective action and physical quantities on the
configurational space metric choice \cite{metric} in the quantum
gravity. Following \cite{VDW}, the configurational space metric is
defined by the highest derivative terms of the classical action.
However, the question on correct choice of the configurational space
metric in the gravity is still open.

The second way is related to the physical parametrization definition.
The best way how the physical parametrization can be defined, we
believe, was suggested by Fujikawa \cite{Fujikawa1}. The anomaly-free
condition for the BRST transformations is assumed to define the
physical variables.  This condition should be imposed on each
variable and means that the tensor densities $\varphi^*$
(obtained when the initial fields $\varphi $ is multiplied by the
corresponding degree of $(-g)$) are thought of as dynamical
variables.  For example, in the quantum gravity either $g_{\mu \nu
}(-g)^{\frac{N-4}{4N}}$ or $g^{\mu \nu }(-g)^{\frac{N+4}{4N}}$ ( $N$
is the space-time dimension) can play the role of a physical dynamical
variable. All matter fields must be replaced with tensor density
fields. These results can also be obtained in the functional integral
approach ~\cite{Fujikawa2} without the BRST-symmetry.

Apart from the parametrization dependence conditioned by
arbitrariness in the dynamical variables choice, an additional
dependence on the parametrization arises due to the quantum field
redefinition at calculations of loop corrections. Splitting the
dynamical variables in the background field method into classical and
quantum parts $\underline{g}^{\ast \mu \nu } = g^{\ast \mu \nu } +
h^{\ast \mu \nu }$, the quantum fields can be replaced with

$$
h^\ast_{\mu \nu } \rightarrow 'h^\ast_{\mu \nu }  = h^\ast_{\mu \nu }
+ \left( h^{\ast 2} \right)_{\mu \nu }
+ \left(h^{\ast 3} \right)_{\mu \nu } + \dots .
$$

\noindent
Such a redefinition has no physical meaning but allows one to
facilitate the calculation of quantum corrections. Parametrization
dependence of a given type is not a subject of the current paper.

In this paper we will analyze the dependence of the one-loop counterterms
on the gauge and parametrization in the Einstein gravity with the
cosmological constant. This paper is organized as follows. In the next
section we calculate the one-loop off-shell counterterms in an arbitrary
gauge and parametrization. Sect.~3 deals with the calculation of the
one-loop counterterms in the special (conformal) parametrization and
in an arbitrary gauge. In sect.~4, the gauge and parametrization
dependence arising in the first-order formalism is investigated.
Sect.~5 concerns the renormalization function calculation off the
mass shell in the quantum gravity. Conclusions present the results.

We use the following notations
$$ c = \hbar = 1;~~~~~ \mu , \nu  = 0,1,2,3;~~~~~ {\it k}^2 = 16 \pi
G,~~~~~(g) = det(g_{\mu \nu }), ~~~~~\varepsilon  = \frac{4-d}{2} $$
$$ R^\sigma _{~\lambda  \mu  \nu } = \partial_\mu \Gamma^\sigma
_{~\lambda \nu }  - \partial_\nu \Gamma^\sigma _{~\lambda \mu }
+ \Gamma^\sigma_{~\alpha \mu } \Gamma^\alpha_{~\lambda  \nu }
- \Gamma^\sigma_{~\alpha  \nu }  \Gamma^\alpha_{~\lambda \mu },~~~~~
R_{\mu \nu } = R^\sigma_{~\mu \sigma \nu },~~~~~
R =  R_{~\mu \nu } g^{\mu \nu }, $$
where $\Gamma^\sigma_{~\mu \nu } $ is the Riemann connection, and
$G$ is the gravitational constant.

Objects marked by the asterisk  $*$ are the tensor densities.
The other variables are the tensors.  Objects marked by tildes are
constructed out of the affine connection $\tilde \Gamma^\sigma_{~\mu
\nu}$. The other objects are the Riemannian ones. Parentheses around
a pair of indices denote the symmetrization whereas parentheses
around four indices mean also the symmetrization with respect to
pairs interchange.

\section{The Einstein gravity in arbitrary gauge and
pa\-ra\-met\-ri\-za\-tion}.

The action of the Einstein gravity with the cosmological constant is

\begin{equation} S = - \frac{1}{{\it k}^2} \int
d^4x   (R-2\Lambda) \sqrt{-g} + \omega \chi ,
\label{action}
\end{equation}

\noindent
where $\omega$ is the dimensionless coupling constant, $\Lambda$ is
the cosmological constant and

\begin{equation}
\chi = \frac{1}{32 \pi^2} \int d^4x \sqrt{-g} \left(
R_{\mu \nu \sigma \alpha} R^{\mu \nu \sigma \alpha}
- 4 R_{\mu \nu} R^{\mu \nu} + R^2 \right),
\label{Euler}
\end{equation}

\noindent
is the Euler number (topological invariant).
Further we will consider the space-time which is topologically
trivial $(\chi = 0)$. Thus, we have the right to use the relation:
$R_{\mu \nu \sigma \alpha} R^{\mu \nu \sigma \alpha}
= 4 R_{\mu \nu} R^{\mu \nu} - R^2 $.
For calculating the one-loop counterterms we use the background field
method \cite{BDW}. We deal also with the invariant regularization
\cite{inv} (dimensional renormalization \cite{dim} and minimal
subtraction scheme \cite{min}). In a general case, either of the
tensor densities $g^\ast_{\mu \nu } = g_{\mu \nu }(-g)^r$ or
$g^{\ast \mu \nu } = g^{\mu \nu}(-g)^p$ is selected as a dynamical
variable of metric theory of the gravity, where $r$ and $p$ are such
numbers that

\begin{equation} {\rm det}
\left| \frac{\partial g^\ast_{\mu \nu }}{\partial g_{\alpha \beta }}
\right| \neq 0, ~~~~ {\rm det} \left| \frac{\partial g^{\ast \mu \nu
}}{\partial g^{\alpha \beta }} \right| \neq 0.
\label{conditions}
\end{equation}

\noindent
In this chapter, under the words first (second) choice of the variables
we understand the using $g^\ast_{\mu \nu } (g^{\ast \mu \nu }) $ as
dynamical variables of the theory. In accordance with the background
field method we rewrite the dynamical field as
$ g^\ast_{\mu \nu } = g^\ast_{\mu \nu } + {\it k} h^\ast_{\mu \nu } $
for the first choice of variables and
$ g^{\ast \mu \nu } = g^{\ast \mu \nu } + {\it k} h^{\ast \mu \nu }$
for the second one.  Here $g^{\ast \mu \nu }$ and $g^\ast_{ \mu \nu }$
are the classical parts which satisfy the following equations

\begin{equation}
\frac{\delta {\it S}_{gr}}{\delta g^{\ast \alpha \beta }}  =
 R_{\alpha \beta}  - \frac{1}{2} R g_{\alpha \beta}
\left(f_1-2f_2 \right) + \Lambda g_{\alpha \beta} f_1   = 0,
\label{variation}
\end{equation}

\noindent
where the values of the coefficients $\{f_i \}$ are given in
Table I.

$$
\begin{array}{|c|c|c|c|c|c|c|} \hline
\multicolumn{7}{|c|}{TABLE ~~~I}             \\   \hline
& f_1 & f_2 & f_3 & f_4 & f_5 & f_6   \\ \hline
g^\ast_{\mu \nu} & \frac{1}{t} & - \frac{r}{t} & 1 & 1 & 0 &
\frac{6 r^2 + 4 r + 1}{t^2} \\ \hline
g^{\ast \mu \nu} &  \frac{1}{s} & \frac{p}{s} & -1 & 0 & 1 &
\frac{6 p^2 - 4p + 1}{s^2} \\ \hline
\end{array}
$$

\noindent
and
\begin{eqnarray}
s & \equiv & 1-4p \neq 0,
\nonumber \\
t & \equiv & 1+4r \neq 0.
\end{eqnarray}

\noindent
Although the Eq.(\ref{variation}) depends on parametrization,
its solution has the standard form:

\begin{equation}
R_{\mu \nu} = \Lambda g_{\mu \nu}.
\label{mass-shell}
\end{equation}

\noindent
In order to calculate quantum corrections, the action (\ref{action})
should be expanded in quantum fields powers. The one-loop diagrams
are generated by the effective action only as quadratic in quantum
fields terms.  The one-loop effective Lagrangian has the following
form:

\begin{eqnarray}
{\it L}_{eff} & = &  \Biggl(
\frac{1}{4} \nabla_\sigma h_{\alpha \beta}
\nabla_\lambda h^{\alpha \beta} g^{\sigma \lambda}
-\frac{1}{2} \nabla_\mu h^\mu_{~\nu} \nabla_\lambda h^{\lambda \nu}
+ \frac{1+f_1}{4} \nabla_\mu h \nabla_\nu h^{\mu \nu}
\nonumber \\
& - & \frac{f_6}{4} \nabla_\sigma h
\nabla_\lambda h g^{\sigma \lambda}  - \frac{1}{2} h^{\mu \nu}
X_{\mu \nu \alpha \beta} h^{\alpha \beta} \Biggr) \sqrt{-g},
\label{effective}
\end{eqnarray}

\noindent
where

\begin{eqnarray}
X_{\mu \nu \alpha \beta} & = &
f_3 R_{\mu \alpha} g_{\nu \beta} + R_{\mu \alpha \nu \beta}
-  \frac{1+f_1}{2} R_{\mu \nu} g_{\alpha \beta}
+ \frac{(1+f_1)^2}{16} R g_{\mu \nu} g_{\alpha \beta}
 \nonumber \\
& - & \frac{(1+f_1)f_3}{4} R g_{\mu \alpha } g_{\nu \beta}
+ f_1 f_3 \Lambda g_{\mu \alpha} g_{\nu \beta}
- \frac{\Lambda}{2} f_1^2 g_{\mu \nu} g_{\alpha \beta},
\label{X}
\end{eqnarray}

\noindent and
$ h \equiv  h_{\alpha \beta } g^{\alpha \beta } $.

The effective Lagrangian (\ref{effective}) is invariant under
general transformations of the coordinates
$x^\mu  \rightarrow  'x^\mu = x^\mu + { k} \xi^\mu(x).$
The quantum fields are then transformed as follows:

\begin{eqnarray}
h^\ast_{ \mu \nu}(x)    \rightarrow   'h^\ast_{ \mu \nu}(x) & = &
h^\ast_{ \mu \nu}(x) - \partial_\mu \xi^\lambda
g^\ast_{ \lambda \nu}(x) - \partial_\nu \xi^\lambda
g^\ast_{ \mu \lambda} \nonumber \\*
& - & \xi^\alpha \partial_\alpha g^\ast_{ \mu \nu}(x)
- 2r \partial_\alpha \xi^\alpha g^\ast_{ \mu \nu}(x) + O({ k}),
\label{coordinatem1}
\end{eqnarray}

\noindent
for the first set of variables and

\begin{eqnarray}
h^{\ast \mu \nu}(x)    \rightarrow   'h^{\ast \mu \nu}(x) & = &
h^{\ast \mu \nu}(x) + \partial_\lambda \xi^\mu
g^{\ast \lambda \nu}(x) + \partial_\lambda \xi^\nu
g^{\ast \mu \lambda}
\nonumber \\*
& - & \xi^\alpha \partial_\alpha g^{\ast \mu \nu}(x)
- 2p \partial_\alpha \xi^\alpha g^{\ast \mu \nu}(x) + O({ k}),
\label{coordinatem2}
\end{eqnarray}

\noindent
for the second choice of variables.

Let us analyze the influence of the gauge and parametrization on the
structure of propagators. We can scoop information on
the particle content of the theory out of the propagator form. For
this, we choose the following gauge:

\begin{equation}
{\it L}_{gf} = \frac{1}{2 \alpha} \left(
\nabla_\mu h^{\ast \mu}_{\nu} - \gamma \nabla_\nu h^{\ast}
\right)^2 \sqrt{-g}.
\end{equation}

\noindent
To obtain the propagator, the metric should be taken as
expansion around the flat background. In this expansion $\Lambda$
should be equal to zero \cite{ChD}. Going to the momentum space and
using the method of projection operators \cite{st,BOS}, we obtain the
following graviton propagator:

\begin{eqnarray}
D_{\mu \nu \alpha \beta}(p^2) & = & \frac{-2 i}{ (2 \pi)^4 p^2}
\Big(P^{(2)}_{\mu \nu \alpha \beta}
+ \alpha P^{(1)}_{\mu \nu \alpha \beta}
+ \frac{b_1}{\Theta} P^{(0-s)}_{\mu \nu \alpha \beta}
+ \frac{b_2}{\Theta} P^{(0-w)}_{\mu \nu \alpha \beta}
\nonumber \\*
& - & \frac{b_3}{\Theta} \left(
P^{(0-sw)}_{\mu \nu \alpha \beta} + P^{(0-ws)}_{\mu \nu \alpha \beta}
\right) \Big),
\label{graviton}
\end{eqnarray}

\noindent
where

\begin{eqnarray}
\Theta & = & - 4 \left(|f_1-f_2| - \gamma \right)^2 \neq 0,
~~~~~b_1 = 2 (1 - \gamma) ^2 - 6 f_2^2 \alpha,
\nonumber \\
b_2 & = & 6 \gamma ^2 - 2 \left(f_1 - f_2 \right)^2 \alpha,
~~~~~~~~~b_3 = \gamma^2 - \gamma - f_2 \left( 1 - f_2 \right) \alpha.
\nonumber
\end{eqnarray}

\noindent
The ghost propagator is given by

\begin{eqnarray}
D_{\mu \nu } (p^2) & = &
\frac{-i}{ (2 \pi)^4 p^2} \Big( \delta_{\mu \nu} -
\frac{ (1+f_1)/4 - \gamma}{(f_1-f_2) - \gamma}
\frac{p_\mu p_\nu}{p^2} \Big).
\end{eqnarray}

\noindent
Expression (\ref{graviton}) describes the propagation of the
following particles: one massless field of spin-two particle named
the graviton, one massless field of spin-one and several massless
scalar fields. Note that the coefficient in front of $P^{(2)}$ in
(\ref{graviton}) does not depend on the choice of the gauge and
parametrization. On the contrary, the coefficients in front of the
other particles are dependent on the gauge and parametrization.
However, it is impossible to choose such a gauge and parametrization
that only $P^{(2)}$ will propagate. In the trivial parametrization
expression (\ref{graviton}) coincides with the result given in
ref.~ \cite{propagator}.

Now we investigate the gauge and parametrization dependence of the
one-loop counterterms. We fix the gauge invariance by the following
condition:

\begin{equation}
{\it L}_{gf} = \frac{1}{2(3+\alpha)} F^\mu F_\mu \sqrt{-g},
\label{gf}
\end{equation}
\begin{equation}
F_\mu =  \nabla_\beta h^{\ast \beta}_{~~\mu}
- \frac{1}{4}\left(1 + f_1 \left( 3 - \frac{2}{\gamma} \right) \right)
\nabla_\mu h^{\ast}.
\label{gauge}
\end{equation}

The ghost action obtained in the standard way is

\begin{equation}
{\it L}_{gh} = \bar c^\alpha \left( g_{\alpha \beta} \nabla^2
- \left( 1- \frac{1}{\gamma} \right) \nabla_\alpha
\nabla_\beta + R_{\alpha \beta} \right) c^\beta \sqrt{-g}.
\label{ghost}
\end{equation}

It should be noted that the calculation of one-loop counterterms in
arbitrary gauge and parametrization is an extremely cumbersome task.
Note that the nonminimal operator of second order arises at calculation.
Although the algorithm developed in \cite{algorithm} can be used for
its calculation, we consider that the simplest way is to use the
common expression for counterterms of arbitrary-order nonminimal
operator given in \cite{our3}. By using the tensor package of
analytic calculations written in REDUCE \cite{tensor}, we obtain that
the one-loop divergences in the gauge (\ref{gf}), (\ref{gauge}) and in
arbitrary parametrization which include the quantum and ghost fields
contributions are the following:

\begin{eqnarray}
\triangle \Gamma^{(1)}_{\infty} & =&  - \frac{1}{32 \pi^2 \varepsilon}
\int d^4x
\Big[ c_1 R_{\mu\nu} R^{\mu \nu} + c_2 \Lambda^2 + c_3 R\Lambda
+ c_4 R^2 \Big] \sqrt{-g},
\label{result}
\end{eqnarray}

\noindent
where $\{c_i \}$ are gauge and parametrization dependent.
On mass-shell defined by eq. (\ref{mass-shell}), we obtain that

\begin{equation}
4 c_1 + c_2 + 4 c_3 + 16 c_4 = - \frac{58}{5},
\label{sum}
\end{equation}

\noindent
or, in the other words,

\begin{equation}
\triangle \Gamma^{(1) On-Shell}_{\infty} =
\frac{1}{32 \pi^2 \varepsilon}
\int d^4x \frac{58}{5} \Lambda^2 \sqrt{-g}.
\label{on-result}
\end{equation}

\noindent
This result coincides with that one obtained in ref.~\cite{ChD}.
In order to present the result in a more compact form, we write
$c_4$ in the Eq.(\ref{result}) as follows:
\begin{equation}
c_4 =
-\frac{1}{4} \left( \frac{29}{10} + c_1 + c_3 + \frac{c_2}{4} \right),
\label{compact}
\end{equation}
the rest coefficients have the following form:

\begin{eqnarray}
c_1 =&&
 - 1/3 f_3 \alpha^2 \gamma^4 - 2 f_3 \alpha \gamma^3 + f_3 \alpha
\gamma^2 - 8/3 f_3\gamma^2 + 2 f_3 \gamma
\nonumber\\&&
+ 4/3 f_5 \alpha^2 \gamma^2 + 2/3 f_5 \alpha^2 + 10/3 f_5\alpha
\gamma^2 + 4 f_5 \alpha \gamma + 8/3 f_5 \alpha
\nonumber\\&&
+ 28/3 f_5 \gamma + 32/3 f_5 + 1/3\alpha^2 \gamma^4 + 1/6 \alpha^2
\gamma^2 + 1/2 \alpha^2 + 2 \alpha \gamma^3
\nonumber\\&&
+ 1/3\alpha \gamma^2 + 7/3 \alpha + 8/3 \gamma^2 - 2 \gamma + 131/30
+ f_1^{-1} \gamma (1/3 f_3 \alpha^2 \gamma^3
\nonumber\\&&
+ 2 f_3 \alpha \gamma^2 - f_3 \alpha \gamma + 8/3 f_3 \gamma - 2 f_3
+ 4/3 f_5 \alpha^2 \gamma^3 + 4/3 f_5 \alpha^2 \gamma
\nonumber\\&&
+ 20/3 f_5 \alpha \gamma^2 + 8/3 f_5 \alpha \gamma + 4/3 f_5 \alpha +
20/3 f_5 \gamma - 1/3 \alpha^2 \gamma^3
\nonumber\\&&
- 1/3 \alpha^2 \gamma - 4/3 \alpha \gamma^2 - 4/3 \alpha \gamma - 2/3
\gamma) + f_1^{-2} \gamma^2 (1/6 \alpha^2 \gamma^2
\nonumber\\&&
+ 1/6 \alpha^2 + 2/3 \alpha \gamma + 2/3 \alpha + 1/3),
\nonumber  \\
c_2 =&&
f_1^2 (9/2 \alpha^2 \gamma^4 + 6 \alpha^2 + 36 \alpha \gamma^3 - 12
 \alpha \gamma^2 + 36 \alpha + 72 \gamma^2 - 48 \gamma + 72)
\nonumber\\&&
+ 6 f_1 (f_3 \alpha^2 \gamma^4 + 2 f_3 \alpha^2 + 10 f_3 \alpha
 \gamma^3 - 4 f_3 \alpha \gamma^2 + 12 f_3 \alpha + 22 f_3 \gamma^2
\nonumber\\&&
- 16 f_3 \gamma + 24 f_3 + 4 f_5 \alpha^2 \gamma^4 + 4 f_5 \alpha^2 +
 28 f_5 \alpha \gamma^3 - 8 f_5 \alpha \gamma^2 + 24 f_5 \alpha
\nonumber\\&&
+ 52 f_5 \gamma^2 - 32 f_5 \gamma + 48 f_5 - 2 \alpha^2 \gamma^4 - 2
 \alpha^2 - 14 \alpha \gamma^3 + 4 \alpha \gamma^2 - 12 \alpha
\nonumber\\&&
- 26 \gamma^2 + 16 \gamma - 24) - 8 f_3 \alpha^2 \gamma^4 - 12 f_3
 \alpha^2 - 72 f_3 \alpha \gamma^3 + 24 f_3 \alpha \gamma^2
\nonumber\\&&
- 72 f_3 \alpha - 144 f_3 \gamma^2 + 96 f_3 \gamma - 144 f_3 - 16 f_5
 \alpha^2 \gamma^4 - 24 f_5\alpha^2 - 144 f_5 \alpha \gamma^3
\nonumber\\&&
+ 48 f_5 \alpha \gamma^2 - 144 f_5 \alpha - 288 f_5\gamma^2 + 192 f_5
 \gamma - 288 f_5 + 13 \alpha^2 \gamma^4 + 12 \alpha^2
\nonumber\\&&
+ 84 \alpha\gamma^3 - 24 \alpha \gamma^2 + 72 \alpha + 156 \gamma^2 -
 96 \gamma + 144 + 2 f_1^{-1} \gamma^2 (f_3 \alpha^2 \gamma^2
\nonumber\\&&
+ 6 f_3 \alpha \gamma + 6 f_3 + 4 f_5 \alpha^2 \gamma^2 + 12 f_5
 \alpha \gamma + 12 f_5 - 2 \alpha^2 \gamma^2 - 6 \alpha \gamma - 6)
\nonumber\\&&
+ 1/2 f_1^{-2} \alpha^2 \gamma^4,
\nonumber \\
c_3 =&&
f_1^2 ( - 9/4 \alpha^2 \gamma^4 - 3 \alpha^2 - 18 \alpha \gamma^3 + 6
 \alpha \gamma^2 - 18 \alpha - 36 \gamma^2 + 24 \gamma - 36)
\nonumber\\&&
+ f_1 ( - 3 f_3 \alpha^2 \gamma^4 - 6 f_3 \alpha^2 - 27 f_3 \alpha
 \gamma^3 + 23/2 f_3 \alpha \gamma^2 - 34 f_3\alpha - 57 f_3 \gamma^2
\nonumber\\&&
+ 46 f_3 \gamma - 70 f_3 - 12 f_5 \alpha^2 \gamma^4 - 12 f_5 \alpha^2
 - 84 f_5 \alpha \gamma^3 + 24 f_5 \alpha \gamma^2 - 72 f_5 \alpha
\nonumber\\&&
- 156 f_5 \gamma^2 + 96 f_5 \gamma - 144 f_5 + 6 \alpha^2 \gamma^4 +
 6 \alpha^2 + 42\alpha \gamma^3 - 12 \alpha \gamma^2 + 36 \alpha
\nonumber\\&&
+ 78 \gamma^2 - 48 \gamma + 72) + 4 f_3 \alpha^2 \gamma^4 + 6 f_3
 \alpha^2 + 32 f_3 \alpha \gamma^3 - 35/3 f_3 \alpha \gamma^2
\nonumber\\&&
+ 34 f_3 \alpha + 62 f_3 \gamma^2 - 46 f_3 \gamma + 70 f_3 + 8 f_5
 \alpha^2 \gamma^4 + 12 f_5 \alpha^2 + 64 f_5 \alpha \gamma^3
\nonumber\\&&
- 70/3 f_5 \alpha \gamma^2 + 68 f_5\alpha + 124 f_5 \gamma^2 - 92 f_5
 \gamma + 140 f_5 - 13/2 \alpha^2 \gamma^4 - 6 \alpha^2
\nonumber\\&&
- 40 \alpha \gamma^3 + 34/3 \alpha \gamma^2 - 34 \alpha - 70 \gamma^2
 + 46 \gamma - 70 + f_1^{-1} \gamma^2 ( - f_3 \alpha^2 \gamma^2
\nonumber\\&&
- 5 f_3 \alpha \gamma + 1/6 f_3\alpha - 5 f_3 - 4 f_5 \alpha^2
 \gamma^2 - 12 f_5 \alpha \gamma - 12 f_5 + 2 \alpha^2 \gamma^2
\nonumber\\&&
+ 6 \alpha \gamma + 6) - 1/4 f_1^{-2} \alpha^2 \gamma^4.
\label{res}
\end{eqnarray}

\noindent
We checked that, in the case  $\Lambda=0$ and in the trivial
parametrization $(r = p = 0)$, these results reproduce the results of
ref.~\cite{algorithm,kallosh,gauge}. In the special gauge, these
results coincide with those of ref.~\cite{MKL}.

Let us trace the computing process which allowed us to get these
results. The expansion (\ref{effective}) has been obtained by means
of a computer calculation. The next step was to put all the terms
including the gauge-fixing ones (\ref{gf}), (\ref{gauge}) into the
computer.  The coefficients $\{c_i\}$ in (\ref{result}) have been
automatically calculated with the algorithm written in \cite{our3}.
The only one problem was to present the obtained data in a compact
form. The particular cases  $(\Lambda = 0, r = p = 0,$ on-shell$)$
have been investigated by using the same computer program. It is
worthy to point out once more that the relation (\ref{sum}) has been
tested by calculating $\{c_i \}$  not introducing them from some
theoretical considerations.

Although the more elegant and powerful methods like BRST, BFV
could be appropriate for theoretical studies, in the current paper
we use the well-studied Faddeev-Popov method due to its
simplicity at plain calculation.

\section{The Einstein gravity in an arbitrary gauge and in the
conformal parametrization}.

The parametrization for the metric field considered in the previous
section does not contain special parametrization where the
conditions (\ref{conditions}) are not fulfilled.
This special parametrization is related to the choice of a
conformal factor for the metric field $ \pi^\ast =
(-g)^{\frac{m}{4}}, ~~m \neq 0 $ and the traceless part
$\psi^\ast_{\mu \nu}  =   g_{\mu \nu} (-g)^{-\frac{1}{4}},~~
det \psi^\ast_{\mu \nu} = 1 $ or
$
\psi^{\ast \mu \nu}  =  g^{\mu \nu} (-g)^{\frac{1}{4}},~~
det \psi^{\ast \mu \nu}  =  1$. The technique of calculations in this
parametrization is equivalent to the one used in the previous
chapter. So, we sketch in only the main points.

In accordance with the background field method we rewrite the
dynamical field as $\pi^\ast = \pi^\ast + {\it k} \tau^\ast $
and
$\psi^\ast_{\mu \nu } = \psi^\ast_{\mu \nu } + {\it k}
\omega^\ast_{\mu \nu }$ for the first set of variables or
$\psi^{\ast \mu \nu } = \psi^{\ast \mu \nu } + {\it k}
\omega^{\ast \mu \nu }$ for second one.
In these expansions $\omega^\ast_{\mu \nu }$ is the traceless variable

\begin{equation}
g^{\mu \nu} \omega^\ast_{\mu \nu}  =  0,
\end{equation}

\noindent  $\psi^\ast_{\mu \nu }$ and $\pi^\ast$ are the
classical parts satisfying the following equations:

\begin{eqnarray}
\frac{\delta {\it S}_{gr}}{\delta \psi^{\ast \alpha \beta}}  & = &
 R_{\alpha \beta}  - \frac{1}{4} R g_{\alpha \beta} = 0,
\nonumber \\
\frac{\delta {\it S}_{gr}}{\delta \pi^\ast}  & = &
 R - 4 \Lambda = 0.
\end{eqnarray}

\noindent
As a result, we obtain once more the expression defined
by eq.~(\ref{mass-shell}) for the old mass shell.

The effective Lagrangian which is quadratic in the quantum fields is

\begin{eqnarray}
{\it L}_{eff} & = &  \Biggl(
\frac{1}{4} \nabla_\sigma \omega_{\alpha \beta}
\nabla_\lambda \omega^{\alpha \beta} g^{\sigma \lambda}
-\frac{1}{2} \nabla_\mu \omega^\mu_{~\nu} \nabla_\lambda \omega^{\lambda \nu}
- \frac{a_1}{m}  \nabla_\mu \tau \nabla_\nu \omega^{\mu \nu}
\nonumber \\
& - & \frac{3}{2m^2}  \nabla_\sigma \tau
\nabla_\lambda \tau g^{\sigma \lambda}  - \frac{1}{2} \omega^{\mu \nu}
X_{\mu \nu \alpha \beta} \omega^{\alpha \beta}   - \frac{1}{2} \tau^2 Z
- \frac{a_1}{m} \tau R_{\mu \nu} \omega^{\mu \nu}
\Biggr) \sqrt{-g},
\label{eff2}
\end{eqnarray}

\noindent
where

\begin{equation}
X_{\mu \nu \alpha \beta}  =  - a_1 R_{\mu \alpha} g_{\nu \beta}
+ R_{\mu \alpha \nu \beta}
+  \frac{a_1}{2} \left( R  - 2 \Lambda \right) g_{\mu \alpha }
g_{\nu \beta},
\end{equation}

\begin{equation}
Z = R \frac{1-m}{m^2} - 4 \Lambda \frac{2-m}{m^2},
\end{equation}

\noindent
and the values of the coefficients $\{a_i \}$ are given in Table II.

$$
\begin{array}{|c|c|c|c|} \hline
\multicolumn{4}{|c|}{TABLE ~~~II}        \\   \hline
 & a_1 & a_2  &  a_3                           \\ \hline
g_{\mu \nu} (-g)^{-\frac{1}{4}} & -1 & 0 & 1 \\ \hline
g^{\mu \nu} (-g)^{\frac{1}{4}}  &  1 & 1 & 0 \\ \hline
\end{array}
$$

\noindent
Now the effective Lagrangian (\ref{eff2}) is invariant under the
following general transformations of the coordinates
$x^\mu  \rightarrow  'x^\mu = x^\mu +
{\it k} \xi^\mu(x)$. The quantum fields are then transformed as
follows:

\begin{equation}
\tau^\ast (x)    \rightarrow   '\tau^\ast(x) =
\tau^\ast (x) - \xi^\alpha \partial_\alpha \pi^\ast(x)
- \frac{m}{2} \partial_\alpha \xi^\alpha \pi^\ast(x)
+ O({\it k}),
\end{equation}

\noindent
and

\begin{eqnarray}
\omega^\ast_{ \mu \nu}(x)  & \rightarrow  &  '\omega^\ast_{\mu
\nu}(x) = \omega^\ast_{ \mu \nu}(x) - \partial_\mu \xi^\lambda
\psi^\ast_{ \lambda \nu}(x) - \partial_\nu \xi^\lambda
\psi^\ast_{ \lambda \mu} \nonumber \\*
&& - \xi^\alpha \partial_\alpha \psi^\ast_{ \mu \nu}(x)
+ \frac{1}{2} \partial_\alpha \xi^\alpha \psi^\ast_{ \mu \nu}(x)
+ O({\it k}),
\end{eqnarray}

\noindent
for the first set of variables and

\begin{eqnarray}
\omega^{\ast \mu \nu}(x)  & \rightarrow &  '\omega^{\ast \mu \nu}(x)
= \omega^{\ast \mu \nu}(x) + \partial_\lambda \xi^\mu \psi^{\ast
\lambda \nu}(x) + \partial_\lambda \xi^\nu \psi^{\ast \mu \lambda}
\nonumber \\* && -  \xi^\alpha \partial_\alpha \psi^{\ast \mu \nu}(x)
- \frac{1}{2} \partial_\alpha \xi^\alpha \psi^{\ast \mu \nu}(x)
+ O({\it k}),
\end{eqnarray}

\noindent
for the second one.

We fix the gauge invariance by the condition (\ref{gf}) where
now $F_\mu$ are :

\begin{equation}
F_\mu  = \nabla_\nu \omega^{\ast \nu}_{~~\mu}
+ \frac{a_1}{m} \left( 1 + 2 \gamma \right) \nabla_\mu \tau^\ast.
\end{equation}

The one-loop counterterms off the mass shell have the form
(\ref{result}). Taking into account the eq. (\ref{compact})
the rest coefficients have the following form:

\begin{eqnarray}
c_1 =&&
1/3 \gamma^4 \alpha^2 (a_1 + 1) + 2 \gamma^3 \alpha (a_1 + 1)
\nonumber\\*&&
+ \gamma^2 (4/3 \alpha^2 a_2 + 1/6 \alpha^2 - \alpha a_1
\nonumber\\*&&
+ 10/3 \alpha a_2 + 1/3 \alpha + 8/3 a_1 + 8/3)
\nonumber\\*&&
+ \gamma (4 \alpha a_2 - 2 a_1 + 28/3 a_2 - 2)
\nonumber\\*&&
+ 2/3 \alpha^2 a_2 + 1/2 \alpha^2 + 8/3 \alpha a_2
\nonumber\\*&&
+ 7/3 \alpha + 32/3 a_2 + 131/30,
\nonumber \\
c_2 =&&
\gamma^4 \alpha^2 (1/2 m^2 + m a_1
\nonumber\\*&&
- 8 m a_2 + 2 m + 2 a_1 + 8 a_2 + 5/2)
\nonumber\\*&&
+ 12 \gamma^3 \alpha ( - 2 m a_2 + m + a_1 + 2 a_2 + 2)
\nonumber\\*&&
+ 12 \gamma^2 ( - 2 m a_2 + m - \alpha + a_1 + 2 a_2 + 5)
\nonumber\\*&&
- 48 \gamma + 6 (\alpha^2 + 6 \alpha + 12),
\nonumber \\
c_3 =&&
\gamma^4 \alpha^2 ( - 1/4 m^2 - 1/2 m a_1
\nonumber\\*&&
+ 4 m a_2 - m - a_1 - 4 a_2 - 5/4)
\nonumber\\*&&
+ \gamma^3 \alpha (12 m a_2 - 5 m - 5 a_1 - 20 a_2 - 10)
\nonumber\\*&&
+ \gamma^2 (1/6 m \alpha + 12 m a_2 - 5 m + 1/6 \alpha a_1
\nonumber\\*&&
+ 2/3 \alpha a_2 + 16/3 \alpha - 5 a_1 - 32a_2 - 22)
\nonumber\\*&&
+ 2 \gamma (2 a_2 + 11) - 3 \alpha^2 - 4 \alpha a_2
\nonumber\\*&&
- 16 \alpha - 4 a_2 - 34,
\label{res2}
\end{eqnarray}

\noindent
where the coefficients $\{a_i \}$ are given in Table II.
On the mass shell we have the same result as in (\ref{on-result}).

\section{The gauge and parametrization dependence of the
Einstein gravity in the first-order formalism}

In this section we analyze the additional dependence on the gauge
and parametrization off the mass shell which can arise in the
Einstein gravity in the first-order formalism at the one-loop level.
We consider the theory with Lagrangian of the Hilbert-Einstein
type without matter fields. In order to construct the theory in the
first-order formalism a number of sets of dynamical variables can be
used:  metric and torsion fields, metric and affine connection, their
tetrad analogies, etc.  The main criteria at the choice of the
connection field arising in the first-order approach is similarity
between the mass shell equations in this approach and in the standard
one.  This agreement ensures the equivalence of these theories
at the tree-loop level \cite{second-class}.

We would like to investigate such an equivalence at the one-loop
level off the mass shell for the metric field and on the mass shell
for the connection field.  We consider the most general case of
variables:  metric $g_{\mu \nu}$ and affine connection
$\tilde{\Gamma}^\sigma_{~\mu\nu}$.  In the previous studies
\cite{first-order} a special case of the connection, which is
symmetric in its indices $\tilde{\Gamma}^\sigma_{~\mu \nu} =
\tilde{\Gamma}^\sigma_{~(\mu \nu)}$, has been selected.  This set of
variables as well as any other set can be obtained from our variables
by the Lagrange multipliers inclusion in the Lagrangian when the
corresponding constraints are provided.  No additional gauge and
parametrization dependence of the one-loop counterterms appears due
to the inclusion of these multipliers.  The action with our variables
has the following form:

\begin{equation}
{\it S}_{gr}  =  - \frac{1}{{\it k}^2} \int d^4 x
\left(\tilde{R} (\tilde{\Gamma}^\sigma_{~\mu \nu} )
- 2 \Lambda \right) \sqrt{-g}.
\label{actionf}
\end{equation}

\noindent
In this theory the affine connection
$\tilde{\Gamma}^{\sigma}_{~\mu \nu}$ as well as either of
the tensor densities $g^\ast_{\mu \nu} = g_{\mu \nu} (-g)^r$
and $ g^{\ast \mu \nu} = g^{\mu \nu} (-g)^p $
(where $r \ne - \frac{1}{4}, p \ne \frac{1}{4}$)
can be thought of as dynamical variables.
The motion equations do not depend on the choice of variables
and have the following form:

\begin{eqnarray}
\frac{\delta S_{gr}}{\delta g^{\ast \alpha \beta}}
& = & \Biggl( {\tilde R}_{(\alpha \beta)}
- \frac{1}{2} {\tilde R}
g_{\alpha \beta} \left( f_1 - 2 f_2 \right)
+ \Lambda g_{\alpha \beta} f_1  \Biggr) f_1 = 0 ,
\nonumber \\
\frac{\delta S_{gr}}{\delta \tilde{\Gamma}^{\sigma}_{~\mu \nu}}
& = &
D^{\alpha \nu}_{~~ \nu} \delta^\beta_\lambda
+ D^\nu_{~\lambda \nu} g^{\alpha \beta}
- D^{\alpha \beta}_{~~ \lambda}
- D^{\beta~\alpha}_{~\lambda} = 0 ,
\end{eqnarray}

\noindent
where
$D^\sigma_{~\mu \nu} =
\tilde{\Gamma}^\sigma_{~\mu \nu}-\Gamma^\sigma_{~\mu \nu} $

This equation can be solved and we obtain the on-shell
condition.

\begin{eqnarray}
D^\sigma_{~\mu \nu }  & = & 0 ,
\nonumber \\*
R_{\mu \nu } & = & \Lambda g_{\mu \nu } .
\end{eqnarray}

Let us analyze in more detail the first set of dynamical variables.
In accordance with the background field method, we
rewrite the dynamical variables as
$ \tilde{\Gamma}^{\sigma}_{~\mu \nu} = \tilde{\Gamma}^{\sigma}_{~\mu \nu}
+ {\it k} \gamma^{\sigma}_{~\mu \nu}$ and
$ g^\ast_{\mu \nu} = g^\ast_{ \mu \nu}  + {\it k} h^\ast_{\mu \nu}
$. The Lagrangian for the calculation of the one-loop counterterms has the
following form:

\begin{eqnarray}
{\it L}_{eff} & = & - \frac{1}{2}
\gamma^\sigma_{~\mu \nu} F_{\sigma~~\lambda}^{~\mu\nu~\alpha\beta}
\gamma^\lambda_{~\alpha \beta}
\nonumber \\
& + &
f_3 \gamma^\lambda_{~\alpha
 \beta} \biggl( B_{\lambda ~~~~\mu \nu}^{~\alpha \beta
 \sigma} \nabla_\sigma - \triangle_{\lambda ~~\mu \nu}^{~\alpha
 \beta} \biggr) \biggl( \frac{(1+f_1)}{4} h g^{\mu \nu} - h^{\mu \nu}
 \biggr)
\nonumber \\
& - & \frac{1}{2} h^{\mu \nu} h^{\alpha \beta}
\Biggl( 2 f_4 R_{\alpha \mu}g_{\beta \nu}
- \frac{(1+f_1)}{2} R_{\mu \nu}  g_{\alpha \beta}
+ f_1 f_3 \Lambda g_{\alpha \mu} g_{\beta \nu}
\nonumber \\
& - &
f_1^2 \frac{\Lambda}{2}  g_{\alpha \beta } g_{\mu \nu}
+ \frac{(1+f_1)^2}{16} R g_{\mu \nu} g_{\alpha
\beta} - \frac{(1+f_1)}{4} f_3 R g_{\alpha \mu} g_{\beta \nu}
\Biggr),
\label{536}
\end{eqnarray}

\noindent
where

\begin{eqnarray}
B_{\lambda~~~\mu \nu}^{~\alpha \beta \sigma} & = &
\delta^\sigma_\lambda \delta^\alpha_\mu \delta^\beta_\nu
- \delta^\beta_\lambda \delta^\alpha_\mu \delta^\sigma_\nu ,
\nonumber \\
F_{\alpha ~~\mu}^{~\beta \lambda~\nu \sigma} & = &
g^{\beta \lambda} \delta^\nu_\alpha \delta^\sigma_\mu
- g^{\beta \sigma} \delta^\nu_\alpha \delta^\lambda_\mu
+ g^{\nu \sigma} \delta^\lambda_\alpha \delta^\beta_\mu
- g^{\lambda \nu} \delta^\sigma_\alpha \delta^\beta_\mu ,
\nonumber \\
\triangle_{\lambda ~~\mu \nu}^{~ \alpha \beta} &  \equiv &
D^\alpha_{~\mu \nu} \delta^\beta_\lambda + D_\lambda
\delta^\alpha_\mu \delta^\beta_\nu - D^\alpha_{~\mu \lambda}
\delta^\beta_\nu - D^\beta_{~\lambda \nu} \delta^\alpha_\mu ,
\nonumber
\end{eqnarray}

\noindent
and the coefficients $\{ f_i \}$ are given in Table I.

It should be noted that the Lagrangian (\ref{actionf}) is invariant
with respect to the extra local projective transformations
\cite{project}

\begin{eqnarray}
g_{\mu  \nu }(x) & \rightarrow & 'g_{\mu \nu }(x)  = g_{\mu \nu }(x),
\nonumber \\*
\Gamma^\sigma _{~\mu \nu }(x) & \rightarrow  &
'\Gamma^\sigma _{~\mu \nu }(x)  = \Gamma^\sigma _{~\mu \nu }(x)
+ \delta^\sigma_\mu C_\nu(x),
\end{eqnarray}

\noindent
where $C_\nu (x)$ is an arbitrary vector.
To define the propagator of the quantum affine connection, one needs
to fix the projective invariance at the quantum level. It was shown
in \cite{MKL2} that the special gauge could be chosen so that the
corresponding projective ghosts do not contribute to the one-loop
divergent terms.  Using this special gauge defined by the following
conditions:

\begin{eqnarray}
f_\lambda & = & \biggl(
B_1g_{\lambda \sigma} g^{\alpha \beta} +
B_2\delta^\alpha_\sigma \delta^\beta_\lambda +
B_3 \delta^\beta_\sigma \delta^\alpha_\lambda
\biggr) \gamma^\sigma_{~\alpha \beta} \equiv
f_{\lambda  \sigma }^{~~~\alpha  \beta }
\gamma^\sigma_{~\alpha  \beta }
\label{prgf}
\\
{\it L}_{gf}  & = & \frac{1}{2} f_\mu f^\mu
\end{eqnarray}

\noindent
where $\{ B_j \}$ are the constants satisfying the only condition:
$B_1 + B_3 + 4B_2 \neq 0 $, we go to new variables by means of the
following replacement:

\begin{equation}
\gamma^\sigma_{~\mu \nu} \rightarrow
\tilde \gamma^\sigma_{~\mu \nu} =
\gamma^\sigma_{~\mu \nu} - \frac{f_3}{2}
\tilde{F}^{-1~\sigma~~\lambda}_{~~~~~\mu \nu~\alpha \beta}
\biggl(B_{\lambda~~~\eta \epsilon}^{~\alpha \beta \tau} \nabla_ \tau
- \triangle_{\lambda ~~\eta \epsilon}^{~\alpha \beta} \biggr)
\biggl( \frac{(1+f_1)}{4} h g^{\eta \epsilon}
 -  h^{\eta \epsilon} \biggr).
\end{equation}

\noindent
where $\tilde{F}^{-1}$ has the following form

\begin{eqnarray}
\tilde{F}^{-1 \alpha ~~\mu}_{~~~~\beta \sigma ~\nu\lambda} & = &
 -\frac{1}{4}  g^{\alpha \mu} g_{\beta \sigma} g_{\nu \lambda} +
  \frac{1}{2}  g^{\alpha \mu} g_{\beta \nu} g_{\sigma \lambda}
- \frac{1}{4} g_{\nu \beta} \delta^\mu_\lambda \delta^\alpha_\sigma
\nonumber \\
& + & \frac{1}{4}  \biggl( g_{\nu \lambda} \delta^\mu_\beta
\delta^\alpha_\sigma
+ g_{\beta \sigma} \delta^\alpha_\nu \delta^\mu_\lambda \biggr)
 - \frac{1}{2} \biggl( g_{\nu \sigma}
\delta^\alpha_\lambda \delta^\mu_\beta + g_{\beta \lambda}
\delta^\mu_\sigma \delta^\alpha_\nu \biggr)
\nonumber \\
& + &
\frac{1}{4} \left( \frac{B_1 - B_3 + 2B_2}{B_1 + B_3 + 4B_2} \right)
\biggl( g_{\nu \lambda}
\delta^\mu_\sigma \delta^\alpha_\beta + g_{\beta \sigma}
\delta^\alpha_\lambda \delta^\mu_\nu \biggr) \nonumber \\
\nonumber \\
& + &
\frac{1}{4} \left( \frac{B_3 - B_1 + 2B_2}{B_1 + B_3 + 4B_2} \right)
\biggl( g_{\beta \lambda}
\delta^\mu_\nu \delta^\alpha_\sigma + g_{\beta \nu} \delta^\alpha_\beta
\delta^\mu_\lambda \biggr) \nonumber \\
\nonumber \\
& + &
\frac{4 - B_1^2 - B_3^2 - 12B_2^2 + 10B_1B_3 - 4B_1B_2 - 4B_2B_3}
{4 (B_1 + B_3 + 4B_2)^2}
g_{\sigma\lambda} \delta^\mu_\nu \delta^\alpha_\beta
\end{eqnarray}

\noindent
The one-loop effective Lagrangian with new variables
on the mass shell for the affine connection has the
following form
($D^\sigma_{~\mu \nu } = 0$)

$$
{\it L}_{eff} =  - \frac{1}{2}
\tilde{\gamma}^\sigma_{~\mu \nu}
\left( F_{\sigma~~\lambda}^{~\mu\nu~\alpha\beta} +
f_{\tau  \sigma }^{~~~\mu \nu} f^{\tau~~~\alpha  \beta }_{~~\lambda}
\right)
\tilde{\gamma}^\lambda_{~\alpha \beta} + { L}_{add}(h),
$$

\noindent
where ${\it L}_{add}(h)$ depends only on the quantum and background
parts of the metric and coincides with the Lagrangian
(\ref{effective}). The calculation results for the affine connection
($D^\sigma_{~\mu \nu } = 0$) on the mass shell and for the metric
field off the mass shell are in excellent agreement with expressions
(\ref{res}) and (\ref{res2}), respectively.

It is easy to show that for the action (\ref{actionf}) two sets of
dynamical variables (as metric and affine connection as tetrad and
affine local Lorentz connection) give the same results at the
one-loop level off the mass shell.

\section{The renormalization group functions cal\-culati\-ons off the
mass shell}

It should be noted once more that the Einstein gravity is
renormalizable at the one-loop level on the mass shell.  All the
one-loop divergences can be removed by only the cosmological constant
renormalization. The renormalization of topological constant is
trivial due to its independence of the gauge and parametrization.
The cosmological constant can be written as

\begin{equation}
\Lambda = \frac{\lambda} {{\it k}^2},
\label{representation}
\end{equation}

\noindent
where $\lambda$ is the dimensionless constant.
Two dimensional parameters can be connected each other through a new
dimensionless constant. As rule, that new constant has no direct physical
meaning, but turns out to be convenient at various (in particular,
renormalization group) calculations. Moreover, in the representation
(\ref{representation}) counterterms are polynoms of dimensionless
coupling constant -- which is in agreement with the standard quantum field
theory \cite{Shirkov}. It is easy to get the renormalization group
equation within this approach

\begin{equation}
\beta_{\lambda} = - \frac{\lambda^2}{16 \pi^2} \frac{29}{10},
\label{beta-on}
\end{equation}

\noindent
where we have used the definition
$\mu^2 \frac{d}{d\mu^2} \lambda =
- \varepsilon \lambda + \beta_\lambda $,
and $\mu ^2$ is the renormalization point.  In this approach there is
asymptotical freedom for the renormalized constant $\lambda$.
The gravitational constant is not renormalized: $(\beta_G = 0)$.
Note that

\begin{equation}
\beta_{\Lambda} = \frac{\beta_{\lambda}}{\lambda} - \beta_G,
\label{cosmological}
\end{equation}

\noindent
where $\mu^2 \frac{d}{d\mu^2} \Lambda = \beta_\Lambda \Lambda.$

Let us consider the other approach. If the theory is on-shell
renormalizable (finite), such a gauge or parametrization should
exist that this theory will also be off-shell renormalizable
(finite) at the one-loop level.  This means that the following
conditions should be imposed on $\{ c_i \} $ (see eqs.
(\ref{result}), (\ref{compact}) ):

\begin{equation}
c_1=0, ~~~~~~ c_3 + \frac{c_2}{4} = -\frac{29}{10}.
\label{ren-cond}
\end{equation}

In this case, there are divergent terms at the one-loop level
($\Lambda^2$ and $\Lambda R$). Supposing that all the divergences can
be absorbed into the gravitational and the cosmological constant, we
obtain the following renormalization group equations:

\begin{eqnarray}
\beta_\lambda  & = &
- \frac{\lambda^2}{16 \pi^2} \frac{29}{10},
\nonumber \\
\beta_G & = &\frac{\lambda}{16 \pi^2} \frac{c_3}{2},
\end{eqnarray}

\noindent
where $\beta_G$ is the $\beta$-function of the gravitational constant G
defined by the relation $\mu^2 \frac{d}{d \mu^2} G = \beta_G G $.
We see that $\beta_\lambda$ is in excellent agreement with the
previous results.  However, the $\beta$-function of the gravitational
constant $G$ behaves differently in these calculation approaches.  In
the second case, $\beta_G$ obviously depends on the gauge and/or
parametrization. If only we think of the Newton constant $G$ as of a
constant with no physical meaning (this constant cannot be observed
in any experiment), then a dependence of $\beta_G$ on gauge
(parametrization) is not dangerous. But the Newtonian potential is
one of the indirectly observable quantities, in our opinion; in
particular, serves for detection of fifth interaction \cite{force}.
So, it is natural that quantum corrections to the Newtonian potential
should not depend on unphysical parameters (gauge and/or
parametrization).  Therefore, the $\beta_G$ has to be gauge and
parametrization independent variable in the MS-scheme and the
dimensional regularization \cite{beta}. The same arguments are
correct for $\beta_{\Lambda}$ also (see eq.(\ref{cosmological})).

To remove this dependence, we suggest that the method of the
renormalization group functions calculation in the gravity should be
modified \cite{MKL-RG}.  Let us analyze the renormalization
procedure in the one-loop approximation with the special gauge in
more detail.  Due to the renormalizability at the one-loop level, all
the one-loop Green functions are renormalizable too. As a
consequence, three renormalization constants may appear:
renormalization constant $Z_G$, renormalization constant $Z_\lambda$
and  renormalization constant $Z_g$ of the metric field
\footnote{The non-zero renormalization constant of the metric field
in the $2+\varepsilon$ gravity has been considered in
ref. \cite{2D,2D2}.} $g_{\mu \nu}$.

The renormalization of the background field naturally arises in the
background field method. Following the arguments of \cite{Abbott}, it
can be shown that the renormalization of quantum and ghost fields in
the background field method is not essential to the one-loop
background Green functions calculation. Since the tensor densities
are thought of as dynamical variables in our calculations, it is
natural to suppose that just these dynamical variables should be
renormalized instead of the metric tensor. All one-loop divergences
can be absorbed into these renormalization constants. We use the
minimal subtraction \cite{min}, so that all the $Z's$ contain only
poles in $\varepsilon$.  Let us consider the parametrization which
has been studied in sect.~2. Supposing that after the multiplicative
redefinition of the metric and parameters

\begin{eqnarray}
g^\ast_{\mu \nu} & \to & g_{\mu \nu}^{\ast B} =
Z_g g_{\mu \nu}^{\ast } =
\left(1 + f_3 \delta Z_g \right) g_{\mu \nu}^{\ast },
\label{field}
\\*
\lambda & \to & \lambda_B = Z_\lambda \lambda =
\left( \lambda + \delta Z_\lambda \right),
\nonumber \\*
G & \to & G_B = Z_G G =
\left( 1 + \delta Z_G \right) G ,
\label{const}
\end{eqnarray}

\noindent
the one-loop background Green functions obtained from the effective
action $\Gamma_R = \Gamma_B^{GR} + \Gamma_{div}^{GR}$ should be
finite as $\varepsilon \to 0$, we have

\begin{eqnarray}
\Gamma_R  & = &
\int d^4x\ \sqrt{-g} \Biggl\{
- \frac{R}{{\it k}^2}
\Big(1+ f_3 f_1 \delta Z_g \Big) \Big(1-\delta Z_G \Big)
\nonumber \\*
&&
+ \frac{2}{{\it k}^4} \Big(\lambda + \delta Z_\lambda \Big)
\Big(1 + 2 f_3 f_1 \delta Z_g \Big) \Big(1 - 2 \delta Z_G \Big)
\nonumber \\*
&&
- \frac{1}{\varepsilon} \frac{1}{16 \pi^2} \biggl[
\frac{c_3}{2} \lambda \frac{R}{{\it k}^2}
- \left( \frac{29}{5} + 2 c_3 \right) \frac{\lambda^2}{{\it k}^4}
\biggr] \Biggr \} ,
\label{ren1}
\end{eqnarray}

\noindent
where we have taken into account that
$$
g^{\mu \nu}_B =  \left(1 - f_3 f_1 \delta Z_g \right) g^{\mu \nu} ,
~~~~~
\sqrt{g_B} = \left(1 + 2 f_3 f_1 \delta Z_g \right) \sqrt{g},
$$

\noindent
and all $\{f_i \}$ are given in Table $I$.
From (\ref{ren1}) we obtain two equations for the renormalization
constant $\delta Z_i$

\begin{eqnarray}
\delta Z_\lambda & = & - \frac{1}{\varepsilon}
\frac{\lambda^2}{16 \pi^2} \frac{29}{10},
\label{system2}
\\
\delta Z_G  - f_3 f_1 \delta Z_g & = & \frac{1}{\varepsilon}
\frac{\lambda}{16 \pi^2} \frac{c_3}{2}.
\label{system}
\end{eqnarray}

\noindent
To find $\delta Z_G$ and $\delta Z_g$, we need one additional equation.
We do not know such an equation (see also \cite{2D} ).
The requirement of on-shell renormalizability where
only physical parameters of the theory are renormalized (using the
on-shell renormalization group equations, in the other words)
is not used in the standard field theory; it is a difficult to extract
information condition in the case of gravity with matter fields.
The fixation of the one-loop $\delta Z_g$, by using the renormalization
group method at the two-loop level \cite{last}, does not work in the
Einstein gravity due to its nonrenormalizability at the two-loop
level. For the time being we do not know how to solve the
Eq.(\ref{system}), but we can show that $\delta Z_g \neq 0$. Let us
rewrite the Eq.(\ref{system}) in the following form:

\begin{equation}
\delta Z_G - \theta \delta Z_g  = \frac{1}{\varepsilon}
\frac{\lambda}{16 \pi^2} \frac{c_3 (\theta)}{2}
\label{new-form}
\end{equation}

\noindent
where the coefficients $\theta$ and $c_3 (\theta)$ depend on choice of
parametrization. To solve eqs.(\ref{ren-cond}), even numerically, is
quite difficult.  But since there is their solution for the trivial
parametrization $(r = p = 0)$ \cite{kallosh}, it is natural to
expect that there is their solution for arbitrary parametrization
also. We will show on example of the conformal parametrization that
$c_3(\theta)$ depends on parametrization for the solutions of
eqs.(\ref{ren-cond}).  The Eq.(\ref{new-form}) written at two
different parametrizations of $\theta_1$ and $\theta_2$ gives:

\begin{equation}
\theta_1 \delta Z_g - \theta_2 \delta Z_g  = \frac{1}{\varepsilon}
\frac{\lambda}{32 \pi^2}
\left[c_3 (\theta_2)  - c_3 (\theta_1) \right]
\label{criterion}
\end{equation}

\noindent
If the constant of metric field renormalization is equal to zero, then
$c_3 (\theta)$ should not depend on parametrization, and
Eq. (\ref{criterion}) is equal to zero. Therefore,
the dependence of $c_3 (\theta)$ on parametrization (not on gauge!)
can be considered as a criterion of the nonzero constant for
metric field renormalization
\footnote{This statement is valid also in the $R^2$-gravity.
The corresponding calculations in arbitrary parametrization
have been carried out in \cite{p-last}.}. It should be noted that
the existence of such a nonzero constant leads to a modification
of anomalous dimensions of matter fields. However, the change of
unphysical renormalization group functions must not influence on
physical observable.

Now, we consider the conformal parametrization mentioned in sect.~3.
In this parametrization  four renormalization constants appear:
renormalization constant $Z_G$,
renormalization constant $Z_\lambda$ ,
renormalization constant $Z_\pi$ of the conformal factor of the metric
field $\pi^\ast$ and renormalization constant Z$_\psi$ of the traceless
field $\psi^\ast_{\mu \nu}$. In this case, expression
$(\ref{field})$ is replaced with the following one:

\begin{eqnarray}
\pi^\ast & \to & \pi^{\ast B} =
Z_\pi \pi^{\ast } =
\left(1 + \delta Z_\pi \right) \pi^{\ast } ,
\nonumber \\*
\psi^\ast_{\mu \nu} & \to & \psi_{\mu \nu}^{\ast B} =
Z_\psi \psi_{\mu \nu}^{\ast } =
\left(1 - a_1 \delta Z_\psi \right) \psi_{\mu \nu}^{\ast } ,
\end{eqnarray}

\noindent
where all $\{a_i \}$ are given in Table $II$. The above arguments and
the relations
$$ g^{\mu \nu}_B =
\left(1 + a_1 \delta Z_\psi - \frac{1}{m} \delta Z_\pi \right) g^{\mu \nu} ,
~~~ \sqrt{g_B} = \left(1 + \frac{2}{m} \delta Z_\pi \right) \sqrt{g}, $$
give the following equations for the renormalization constants:

\begin{eqnarray}
\delta Z_\lambda + 2 \lambda \left(
\frac{1}{m} \delta Z_\pi - \delta Z_G \right)
& = & - \frac{1}{\varepsilon} \frac{\lambda^2}{32 \pi^2}
\left( \frac{29}{5} + 2 c_3 \right) ,
\nonumber \\*
\delta Z_G  - a_1 \delta Z_\psi - \frac{1}{m} \delta Z_\pi
& = & \frac{1}{\varepsilon} \frac{\lambda}{32 \pi^2} c_3 ,
\label{system1}
\end{eqnarray}

\noindent
which are conditioned by the renormalizability at the one-loop level
off the mass shell. Let us rewrite these equations as:

\begin{eqnarray}
\delta Z_\lambda & = &
- \frac{1}{\varepsilon} \frac{\lambda^2}{16 \pi^2} \frac{29}{10}
+ 2 \lambda a_1 \delta Z_\psi,
\label{invariant}
\\*
\delta Z_G - \frac{1}{m} \delta Z_\pi & = & \frac{1}{\varepsilon}
\frac{\lambda}{16 \pi^2} \frac{c_3}{2} + a_1 \delta Z_\psi,
\label{invariant2}
\end{eqnarray}

\noindent
It is easily seen that the appearance of one additional renormalization
constant $\delta Z_\psi$ leads to even more complicated solution of
the renormalization group equations. In this situation, the quantum field
theory formalism turns out to be very useful. The one-loop
$\beta$-function, in the MS-scheme and the dimensional regularization,
is gauge, parametrization and scheme independent. So,
$\delta Z_\lambda$  obtained from (\ref{invariant}) should coincide with
$\delta Z_\lambda$ obtained from (\ref{system2})
\footnote{In the one-loop approach
$\beta_\lambda = \left( \lambda \frac{\partial}{\partial \lambda} -
1\right) \delta Z_\lambda,$
$\beta_G = \lambda \frac{\partial}{\partial \lambda} \delta Z_G. $}.
Therefore, we have Eq.(\ref{system2}) at hand and

\begin{eqnarray}
\delta Z_\psi & = & 0,
\label{traceless}
\\*
\delta Z_G - \frac{1}{m} \delta Z_\pi & = & \frac{1}{\varepsilon}
\frac{\lambda}{16 \pi^2} \frac{c_3}{2}
\label{trace}
\end{eqnarray}

\noindent
Eq.(\ref{trace}) has the same disadvantage as Eq.(\ref{system}).
Using the same argumentation as previously, it can be easily shown that
the dependence of $c_3(\theta)$ on parametrization for the solutions of
eqs.(\ref{ren-cond}) is the condition that  $\delta Z_\pi \neq 0.$
A number of numerical solutions of eqs.(\ref{ren-cond}), $c_3$ values,
and the accuracy of calculations are given in Table III (these
results concern the conformal parametrization).

$$
\begin{array}{|c|c|c|c|c|} \hline
\multicolumn{5}{|c|}{TABLE ~~~III}             \\   \hline \hline
\multicolumn{5}{|c|}{a_1 = -1}       \\
\multicolumn{5}{|c|}{a_2 = 0}       \\ \hline \hline
m & -3 & -4 & -5 & -11
\\ \hline
\alpha & -2,042785458 & -1,827706813 & -1,448862095 & -0,4704315204
\\ \hline
\gamma & -2,266606921 & -2,363794178 & -2,609094769 & -4,529683566
\\ \hline
c_1 & 1,05 \times 10^{-9} & 4,54 \times 10^{-9} & 4,26 \times 10^{-10}
& -3,86\times 10^{-8}
\\ \hline
c_3 + \frac{c_2}{4} + \frac{29}{10} & 1,05 \times 10^{-9}
& 1,71 \times 10^{-8} & -9,25 \times 10^{-9} & -4,6 \times 10^{-6}
\\ \hline
c_3 &  4,78451 & 31,4664 & 62,76 & 477,634
\\ \hline
\end{array}
$$

\noindent
As it is easily seen, $c_3$ depends on parametrization, and $\delta
Z_\pi (\delta Z_g) \neq 0.$ The one additional interesting feature
follows the equation (\ref{traceless}):  multiplicative
renormalization of the metric field is related to only the conformal
mode.

So, we showed that at quantization of gravitational field, a nonzero
constant of the metric renormalization appears without fail,
if only the Newtonian constant G (or the cosmological
constant $\Lambda$) is a physical parameter. Then $\delta
Z_G (\beta_G)$ is gauge and parametrization independent quantity,
so that only unphysical parameter - the anomalous dimension
of field - depends on gauge and parametrization. The whole
renormalization group analysis was performed on the basis of
methods and results of the standard quantum field theory.
We were not aimed at introduction of new physical conceptions,
our purpose was to explain the available results from point of view
of the quantum field theory. In particular, for similar reasons
we do not use the on-shell renormalization group equations
(where $\beta_G = 0$, and eqs. (\ref{system}), (\ref{invariant2})
simply define the renormalized constant of the metric). The question
of a physical meaning of the metric field renormalization is also
beyond this paper. At quantization of gravitational field, the metric
as quantum field looses its original physical meaning, so it is
necessary to define some additional relation (renormalization
condition) between renormalized and physical metric.

\section{Conclusions}

In this paper, the detailed analysis of the dependence of the
Einstein gravity with the cosmological constant  in the one-loop
approximation has been carried out.  In particular, the following
questions were considered:

\begin{enumerate}

\item
By using the technique of spin-projective operators, the dependence
of the gravitational field propagator on the gauge and
parametrization has been analyzed.

\item
The one-loop counterterms off the mass shell in an ar\-bit\-ra\-ry
gauge and pa\-ra\-met\-ri\-za\-ti\-on have been calculated (\ref{res},
\ref{res2}).  It was shown that the freedom in the gauge and
parametrization choice (three free parameters) resulted in
the different forms for the counterterms off the mass shell
(arbitrary values of the coefficients $\{c_i \}$).

\item
The dependence of the one-loop counterterms on the gauge and
parametrization in the first-order formalism has been studied. No
additional dependence of the one-loop counterterms off the mass shell
on the equations of the connection motion arisen.

\item
Using as an example the Einstein gravity we have showed that to
perform the renormalization group function calculation off the mass
shell in the $4D$ quantum gravity correctly, one should take into
account the metric field renormalization.  Only in this case the
renormalization group functions will answer a number of the standard
field theory points.  In particular, in the dimensional
regularization \cite{dim} and in the MS-scheme \cite{min} the
one-loop $\beta$-functions are independent of the gauge and
parametrization; all the dependence on unphysical parameters enters
the renormalization constants of the wave functions.

\end{enumerate}

\vspace{1cm}
\noindent
{\bf Acknowledgement}

M.~Yu.~Kalmykov is grateful very much to L.~Avdeev,
J.~Buchbinder, D.~Kazakov and I.~Shapiro for the fruitful
discussions and useful remarks.  P.~I.~Pronin and K.~V.~Stepanyantz
are very grateful to our colleges at the Theoretical Physics
Department.  The authors are grateful to M.~Tentyukov for help in
numerical simulations of eqs. (\ref{ren-cond}). We are indebted to
the referees for the useful comments which allowed us to improve
significantly this paper.  Authors would like to thank
G.~Sandukovskaya for help in editing the text.

\section*{Appendix A}
In this appendix we present several functions of $g_{\mu \nu }$ and
their expansion in powers of the quantum field.
In an arbitrary parametrization considered in sect.~2 we have

\begin{eqnarray}
(-g) & = & (- g^\ast)^{f_1} ,
\nonumber \\
g_{\mu \nu } & = & g^\ast_{\mu \nu }  (- g^\ast)^{f_2} ,
\nonumber \\
g^{\mu \nu } & = & g^{\ast \mu \nu } (- g^\ast)^{-f_2} ,
\nonumber \\
\Gamma^\sigma_{~\mu \nu} & = & \frac{1}{2} g^{\ast \sigma \lambda}
\biggl( - \partial_\lambda g^\ast_{\mu \nu}  +
\partial_\mu g^\ast_{\lambda \nu} +
\partial_\nu g^\ast_{\mu \lambda} \biggr)
\nonumber \\
& + & \frac{f_2}{2}  g^{\ast \alpha \beta}
\partial_\lambda g^\ast_{\alpha \beta}
\biggl( \delta^\sigma_\mu \delta^\lambda_\nu
+ \delta^\sigma_\nu \delta^\lambda_\mu
- g^{\ast \sigma \lambda} g^\ast_{\mu \nu} \biggr) ,
\nonumber \\
(-g)^l & = & (-g)^l \left( 1 + {\it k} l f_1 f_3 h +
 \frac{{\it k}^2}{2}\left( l^2 f_1^2 h^2  - l f_1 f_3 h_{\alpha
\beta } h^{\alpha \beta } \right)  + O({\it k}^3) \right) ,
\nonumber \\
g^{\mu \nu } & = & g^{\mu \nu } - {\it k} f_3
\left( f_2 h g^{\mu \nu }  + h^{\mu \nu } \right) +
{\it k}^2 f_2 h h^{\mu \nu}
\nonumber \\
& + & \frac{{\it k}^2}{2} \left( f_2^2 h^2  + f_2 f_3 h_{\alpha \beta }
h^{\alpha \beta } \right) g^{\mu \nu } + f_4 {\it k}^2 h^{\mu \alpha}
h^\nu_{~\alpha} + O({\it k}^3) ,
\nonumber \\
g_{\mu \nu } & = & g_{\mu \nu } + {\it k} f_3
\left( f_2 h g_{\mu \nu }  + h_{\mu \nu } \right) +
 \frac{{\it k}^2}{2} \left( f_2^2 h^2  - f_2 f_3 h_{\alpha
\beta } h^{\alpha \beta } \right) g_{\mu \nu }
\nonumber \\
& + & {\it k}^2 f_2 h h_{\mu \nu }
+ {\it k}^2 f_5 h_{\mu \alpha} h_\nu^{~ \alpha} + O({\it k}^3) ,
\nonumber \\
\Gamma^\sigma_{~ \mu\nu} & = & \Gamma^\sigma_{~ \mu \nu} -
 f_3 \frac{{\it k}}{2} g^{\sigma \lambda}
 \biggl( \nabla_\lambda h_{\mu \nu}
    - \nabla_\mu h_{\nu \lambda} - \nabla_\nu h_{\mu \lambda} \biggr)
\nonumber \\
&  -  &  \frac{{\it k}}{2} f_2 f_3
g^{\alpha \beta} \nabla_\lambda h_{\alpha \beta}
\biggl(g^{\sigma \lambda} g_{\mu \nu}
- \delta^\lambda_\mu \delta^\sigma_\nu - \delta^\lambda_\nu
\delta^\sigma_\mu \biggr)
\nonumber \\
& + & \frac{{\it k}^2}{2} h^{\lambda \sigma } \left(
\nabla_\lambda h_{\mu \nu }  - \nabla_\mu h_{\nu \lambda } -
\nabla_\nu h_{\mu \lambda }  \right)
\nonumber \\
& - & \frac{{\it k}^2}{2} f_2 f_3
h^{\alpha \beta } \nabla_\lambda
h_{\alpha \beta } \left(\delta^\sigma_\mu  \delta^\lambda_\nu
+ \delta^\sigma_\nu  \delta^\lambda_\mu   -
g^{\sigma \lambda } g_{\mu \nu } \right)
\nonumber \\
& + &
 f_5 \frac{{\it k}^2}{2} g^{\sigma \lambda}
 \biggl(\nabla_\mu \left(h_{\nu \alpha} h_\lambda^{~\alpha} \right)
      + \nabla_\nu \left(h_{\mu \alpha} h_\lambda^{~\alpha} \right)
      - \nabla_\lambda \left( h_{\mu \alpha} h_\nu^{~\alpha} \right)
 \biggr)
\nonumber \\
& + &
 f_2 f_5 \frac{{\it k}^2}{2}
 \biggl( h^{\sigma \lambda} g_{\mu \nu} \nabla_\lambda h
- h_{\mu \nu} g^{\mu \nu } \nabla_\lambda h \biggr) + O({\it k}^3),
\nonumber
\end{eqnarray}

\noindent
where the coefficients $\{ f_i \}$ are given in Table I.
In the conformal parametrization investigated in sect.~3 the expansion
in powers of the quantum field are given below

\begin{eqnarray}
(-g) & = & (\pi^\ast)^{\frac{4}{m}} ,
\nonumber \\
g_{\mu \nu } & = & \psi^\ast_{\mu \nu }  (\pi^\ast)^{\frac{1}{m}} ,
\nonumber \\
g^{\mu \nu } & = & \psi^{\ast \mu \nu } (\pi^\ast)^{-\frac{1}{m}}  ,
\nonumber \\
\Gamma^\sigma_{~\mu \nu} & = & \frac{1}{2} \psi^{\ast \sigma \lambda}
\biggl( - \partial_\lambda \psi^\ast_{\mu \nu}  +
\partial_\mu \psi^\ast_{\lambda \nu} +
\partial_\nu \psi^\ast_{\mu \lambda} \biggr)
\nonumber \\
& + & \frac{1}{2m \pi^\ast}
\partial_\lambda \pi^\ast
\biggl( \delta^\sigma_\mu \delta^\lambda_\nu
+ \delta^\sigma_\nu \delta^\lambda_\mu
- \psi^{\ast \sigma \lambda} \psi^\ast_{\mu \nu} \biggr) ,
\nonumber \\
(-g)^l & = & (-g)^l \left( 1 + {\it k} \frac{4l}{m}\tau +
 {\it k}^2\frac{2l(4l-m)}{m^2}\tau^2
+ {\it k}^2 a_1 \frac{l}{2} \omega_{\mu \nu} \omega^{\mu \nu}
+ O({\it k}^3) \right) ,
\nonumber \\
g^{\mu \nu } & = & g^{\mu \nu } - {\it k}
\left( \frac{1}{m} \tau g^{\mu \nu }  - a_1 \omega^{\mu \nu } \right) +
{\it k}^2 a_3 \omega^\nu_\alpha \omega^{\mu \alpha}
\nonumber \\
& + & {\it k}^2 \left( \frac{1+m}{2 m^2} \tau^2  g^{\mu \nu}
- a_1 \frac{1}{m} \omega^{\mu \nu} \tau \right) + O({\it k}^3) ,
\nonumber \\
g_{\mu \nu } & = & g_{\mu \nu } + {\it k}
\left( \frac{1}{m} \tau g_{\mu \nu }  - a_1 \omega_{\mu \nu } \right) +
 {\it k}^2 \left( \frac{1-m}{2 m^2} \tau^2  g_{\mu \nu}  -
\frac{1}{m} a_1 \omega_{\mu \nu} \tau \right)
\nonumber \\
& + & {\it k}^2 a_2 \omega_{\mu \alpha} \omega_\nu^{~\alpha}
+ O({\it k}^3) ,
\nonumber \\
\Gamma^\sigma_{~ \mu\nu} & = & \Gamma^\sigma_{~ \mu \nu} -
 \frac{{\it k}}{2} a_1 g^{\sigma \lambda}
 \biggl(- \nabla_\lambda \omega_{\mu \nu}
    + \nabla_\mu \omega_{\nu \lambda} + \nabla_\nu \omega_{\mu \lambda} \biggr)
\nonumber \\
&  +  &  \frac{{\it k}}{2m}
 \nabla_\lambda \tau \biggl(-g^{\sigma \lambda} g_{\mu \nu}
+ \delta^\lambda_\mu \delta^\sigma_\nu + \delta^\lambda_\nu
\delta^\sigma_\mu \biggr)
+  \frac{{\it k}^2}{2} \omega^{\lambda \sigma } \left(
\nabla_\lambda \omega_{\mu \nu }  - \nabla_\mu \omega_{\nu \lambda } -
\nabla_\nu \omega_{\mu \lambda }  \right)
\nonumber \\
& - & \frac{{\it k}^2}{2m} \tau \nabla_\lambda \tau
 \left(\delta^\sigma_\mu  \delta^\lambda_\nu
+ \delta^\sigma_\nu  \delta^\lambda_\mu   -
g^{\sigma \lambda } g_{\mu \nu } \right)
-
 \frac{{\it k}^2}{2m} a_1
 \biggl( \omega^{\sigma \lambda} g_{\mu \nu} \nabla_\lambda \tau
 - g^{\sigma \lambda} \omega_{\mu \nu} \nabla_\lambda \tau \biggr)
\nonumber \\
& + &
 \frac{{\it k}^2}{2} a_2
 g^{\sigma \lambda} \biggl(
- \nabla_\lambda \left( \omega_{\mu \alpha} \omega_\nu^{~ \alpha} \right)
+ \nabla_\mu \left( \omega_{\lambda \alpha} \omega_\nu^{~ \alpha} \right)
+ \nabla_\nu \left( \omega_{\mu \alpha} \omega_\lambda^{~ \alpha} \right)
\biggr),
\nonumber
\end{eqnarray}

\noindent
where the coefficients $\{a_i \}$  are given in Table II.

\end{document}